\documentclass[onecolumn,showpacs,prc,floatfix]{revtex4}
\usepackage{graphicx}
\usepackage{amsmath}
\newcommand{\p}[1]{(\ref{#1})}
\newcommand\beq{\begin{eqnarray}} \newcommand\eeq{\end{eqnarray}}
\newcommand\beqstar{\begin{eqnarray*}} \newcommand\eeqstar{\end{eqnarray*}}
\newcommand{\beqe}{\begin{equation}} \newcommand{\eeqe}{\end{equation}}
\newcommand{\beqestar}{\begin{equation*}} \newcommand{\eeqestar}{\end{equation*}}
\newcommand{\bal}{\begin{align}}
\newcommand{\tbf}{\textbf}

\begin {document}
\title{Spin polarized states   in
nuclear matter\protect\\ with Skyrme effective interaction }
\author{ A. A. Isayev}
\affiliation{Kharkov Institute of Physics and Technology,
Academicheskaya Str. 1,
 Kharkov, 61108, Ukraine}
 \author{J. Yang}
 \affiliation{Dept. of Physics and Center for Space Science and Technology,
Ewha Womans University, Seoul 120-750, Korea\\ and  Center for
High Energy Physics, Kyungbook National University, Daegu 702-701,
Korea}
\begin{abstract}  The possibility of appearance of spin polarized states
in symmetric and strongly asymmetric nuclear matter is analyzed
within the framework of a Fermi liquid theory with the Skyrme
effective interaction. The zero temperature dependence of the
neutron and proton spin polarization parameters as functions of
density is found for SkM$^*$, SGII (symmetric case) and SLy4, SLy5
(strongly asymmetric case) effective forces. By comparing free
energy densities, it is shown that in symmetric nuclear matter
ferromagnetic spin state (parallel orientation of neutron and
proton spins) is more preferable than antiferromagnetic one
(antiparallel orientation of spins). Strongly asymmetric nuclear
matter undergoes at some critical density a phase transition to
the state with the oppositely directed spins of neutrons and
protons while the state with the same direction of spins does not
appear. In comparison with neutron matter, even small admixture of
protons strongly decreases the threshold density of spin
instability. It is clarified that protons become totally polarized
within a very narrow density domain while  the density profile of
the neutron spin polarization parameter is characterized by the
appearance of long tails near the transition density.
\end{abstract}
 \maketitle

\section{Introduction}
 The spontaneous appearance of  spin polarized states in nuclear
matter is the topic of a great current interest due to relevance
in astrophysics. In particular, the effects of spin correlations
in the medium strongly influence the neutrino cross section and
neutrino luminosity. Hence, depending on whether nuclear matter is
spin polarized or not, drastically different scenarios of
supernova explosion and cooling of neutron stars can be realized.
Another aspect relates to pulsars, which are considered to be
rapidly rotating neutron stars, surrounded by strong magnetic
field. There is still no general consensus regarding the mechanism
to generate such strong magnetic field of a neutron star. One of
the hypotheses is that magnetic field can be produced by a
spontaneous ordering of spins in the dense stellar core.

 The
possibility of a phase transition of normal nuclear matter to the
ferromagnetic state was studied by many authors. In  the gas model
of hard spheres,  neutron matter becomes ferromagnetic at
$\varrho\approx0.41\,\mbox{fm}^{-3}$~\cite{R}. It was found in
Refs.~\cite{S,O} that the inclusion of long--range attraction
significantly increases the ferromagnetic transition density
(e.g., up to $\varrho\approx2.3\,\mbox{fm}^{-3}$ in the Brueckner
theory with a simple central potential and hard core only for
singlet spin states~\cite{O}). By determining magnetic
susceptibility with Skyrme effective forces, it was shown in
Ref.~\cite{VNB} that the ferromagnetic transition occurs at
$\varrho\approx0.18$--$0.26\,\mbox{fm}^{-3}$. The Fermi liquid
criterion for the ferromagnetic instability in neutron matter with
the Skyrme interaction is reached at
$\varrho\approx2$--$4\varrho_0$~\cite{RPLP}, where
$\varrho_0=0.16\,\mbox{fm}^{-3}$ is the nuclear matter saturation
density.    The general conditions on the parameters of
neutron--neutron interaction, which  result in a magnetically
ordered state of neutron matter, were formulated in
Ref.~\cite{ALP}. Spin correlations in dense neutron matter were
studied within the relativistic Dirac--Hartree--Fock approach with
the effective nucleon--meson Lagrangian in Ref.~\cite{MNQN},
predicting the ferromagnetic transition at several times nuclear
matter saturation density. The importance of the Fock exchange
term in the relativistic mean--field approach for the occurrence
of ferromagnetism in nuclear matter was established in
Ref.~\cite{TT}. The stability of strongly asymmetric nuclear
matter with respect to spin fluctuations was investigated in
Ref.~\cite{KW}, where it was shown that  the system with localized
protons can develop a spontaneous polarization, if  the
neutron--proton spin interaction exceeds some threshold value.
This conclusion was confirmed also by calculations within the
relativistic Dirac--Hartree--Fock approach to strongly asymmetric
nuclear matter~\cite{BMNQ}.

For the models with realistic nucleon--nucleon (NN)
interaction, the ferromagnetic phase transition seems to be
suppressed up to densities well above
$\varrho_0$~\cite{PGS}--\cite{H}. In particular, no evidence of
ferromagnetic instability has been found in recent studies of
neutron matter~\cite{VPR} and asymmetric nuclear matter~\cite{VB}
within the Brueckner--Hartree--Fock approximation with realistic
Nijmegen II, Reid93 and Nijmegen NSC97e NN interactions. The same
conclusion was obtained in Ref.~\cite{FSS}, where magnetic
susceptibility of neutron matter was calculated with the use of
the Argonne $v_{18}$ two--body potential and Urbana IX three--body
potential.

Here we  continue the study of spin polarizability of nuclear
matter with the use of an effective NN interaction. As a framework
of consideration, a Fermi liquid (FL) description of nuclear
matter is chosen~\cite{AKP,AIP}. As a potential of NN interaction,
we use the Skyrme effective interaction, utilized earlier in a
number of contexts for nuclear matter
calculations~\cite{SYK}--\cite{AAI}.  We explore the possibility
of FM and
 AFM phase transitions in nuclear matter, when
the spins of protons and neutrons are aligned in the same
direction or in the opposite direction, respectively. In contrast
to the approach, based on the calculation of magnetic
susceptibility, we obtain the self--consistent equations for the
FM and  AFM spin order parameters and find their solutions at zero
temperature. This allows us to determine not only the critical
density of instability with respect to spin fluctuations, but also
to establish the density dependence of the order parameters and to
clarify the question of thermodynamic stability of various phases.
The main emphasis in our study will be laid on the region of zero
isospin asymmetry (symmetric nuclear matter) and large isospin
asymmetry  (strongly asymmetric nuclear matter and neutron
matter).

Note that we consider  the thermodynamic properties of spin
polarized states in nuclear matter up to the high density region
relevant for astrophysics. Nevertheless, we take into account the
nucleon degrees of freedom only,  although other degrees of
freedom, such as pions, hyperons, kaons,  or quarks could be
important at such high densities.
\section{Basic Equations}
 The normal states of nuclear matter are described
  by the normal distribution function of nucleons $f_{\kappa_1\kappa_2}=\mbox{Tr}\,\varrho
  a^+_{\kappa_2}a_{\kappa_1}$, where
$\kappa\equiv({\bf{p}},\sigma,\tau)$, ${\bf p}$ is momentum,
$\sigma(\tau)$ is the projection of spin (isospin) on the third
axis, and $\varrho$ is the density matrix of the
system~\cite{I,IY}.  The energy of the system is specified as a
functional of the distribution function $f$, $E=E(f)$, and
determines the single particle energy
 \begin{eqnarray}
\varepsilon_{\kappa_1\kappa_2}(f)=\frac{\partial E(f)}{\partial
f_{\kappa_2\kappa_1}}. \label{1} \end{eqnarray} The
self-consistent matrix equation for determining the distribution
function $f$ follows from the minimum condition of the
thermodynamic potential and is
  \begin{eqnarray}
 f=\left\{\mbox{exp}(Y_0\varepsilon+
Y_4)+1\right\}^{-1}\equiv
\left\{\mbox{exp}(Y_0\xi)+1\right\}^{-1}.\label{2}\end{eqnarray}
Here the quantities $\varepsilon$ and $Y_4$ are matrices in the
space of $\kappa$ variables, with
$Y_{4\kappa_1\kappa_2}=Y_{4\tau_1}\delta_{\kappa_1\kappa_2}$
$(\tau_1=n,p)$, $Y_0=1/T,\ Y_{4n}=-\mu_n^0/T$ and
$Y_{4p}=-\mu_p^0/T$ being
 the Lagrange multipliers, $\mu_n^0$ and $\mu_p^0$  the chemical
potentials of  neutrons and protons, and $T$  the temperature.
Since it is assumed to consider a nuclear system with an excess of
neutrons, the positive isospin projection is assigned
 to neutrons.
 Further we shall study the possibility of
formation of various types of spin ordering (with parallel and
antiparallel orientation of neutron and proton spins)
 in nuclear matter.

The normal  distribution function can be expanded in the Pauli
matrices $\sigma_i$ and $\tau_k$ in spin and isospin
spaces
\begin{align} f({\bf p})&= f_{00}({\bf
p})\sigma_0\tau_0+f_{30}({\bf p})\sigma_3\tau_0\label{7.2}\\
&\quad + f_{03}({\bf p})\sigma_0\tau_3+f_{33}({\bf
p})\sigma_3\tau_3. \nonumber
\end{align}
 For the energy functional invariant with
respect to rotations in spin and isospin spaces, the structure of
the single particle energy  is  similar to the structure of the
distribution function $f$: \begin{align} \varepsilon({\bf p})&=
\varepsilon_{00}({\bf
p})\sigma_0\tau_0+\varepsilon_{30}({\bf p})\sigma_3\tau_0\label{7.3}\\
&\quad + \varepsilon_{03}({\bf
p})\sigma_0\tau_3+\varepsilon_{33}({\bf p})\sigma_3\tau_3.
\nonumber
\end{align}
Using Eqs.~\p{2} and \p{7.3}, one can express evidently the
distribution functions $f_{00},f_{30},f_{03},f_{33}$
 in
terms of the quantities $\varepsilon$: \begin{align}
f_{00}&=\frac{1}{4}\{n(\omega_{+,+})+n(\omega_{+,-})+n(\omega_{-,+})+n(\omega_{-,-})
\},\nonumber
 \\
f_{30}&=\frac{1}{4}\{n(\omega_{+,+})+n(\omega_{+,-})-n(\omega_{-,+})-n(\omega_{-,-})
\},\nonumber\\
f_{03}&=\frac{1}{4}\{n(\omega_{+,+})-n(\omega_{+,-})+n(\omega_{-,+})-n(\omega_{-,-})
\},\nonumber\\
f_{33}&=\frac{1}{4}\{n(\omega_{+,+})-n(\omega_{+,-})-n(\omega_{-,+})+n(\omega_{-,-})
\}.\label{2.4}
 \end{align} Here $n(\omega)=\{\exp(Y_0\omega)+1\}^{-1}$ and
\begin{gather*}
\omega_{+,+}=\xi_{00}+\xi_{30}+\xi_{03}+\xi_{33},\;\\
\omega_{+,-}=\xi_{00}+\xi_{30}-\xi_{03}-\xi_{33},\;\\
\omega_{-,+}=\xi_{00}-\xi_{30}+\xi_{03}-\xi_{33},\;\\
\omega_{-,-}=\xi_{00}-\xi_{30}-\xi_{03}+\xi_{33},\;\end{gather*}
where \begin{align*}\xi_{00}&=\varepsilon_{00}-\mu_{00}^0,\;
\xi_{30}=\varepsilon_{30},\;
\\
\xi_{03}&=\varepsilon_{03}-\mu_{03}^0,\;\xi_{33}=\varepsilon_{33},\\
\mu_{00}^0&={\frac{\mu_n^0+\mu_p^0}{2}},\quad
\mu_{03}^0={\frac{\mu_n^0-\mu_p^0}{2}}.\end{align*}
 As follows from the structure of the distribution
functions $f$, the quantity $\omega_{\pm,\pm}$, being the exponent
in the Fermi distribution function $n$, plays the role of the
quasiparticle spectrum. We consider the case when the spectrum is
fourfold split due to the spin and isospin dependence of the
single particle energy $\varepsilon({\bf p})$ in Eq.~\p{7.3}. The
branches $\omega_{\pm,+}$ correspond to neutrons with spin up and
spin down, and  the branches $\omega_{\pm,-}$  correspond to
protons with spin up and spin down.

The distribution functions $f$ should satisfy the norma\-lization
conditions
\begin{align} \frac{4}{\cal
V}\sum_{\bf p}f_{00}({\bf p})&=\varrho,\label{3.1}\\
\frac{4}{\cal V}\sum_{\bf p}f_{03}({\bf
p})&=\varrho_n-\varrho_p\equiv\alpha\varrho,\label{3.3}\\
\frac{4}{\cal V}\sum_{\bf p}f_{30}({\bf
p})&=\varrho_\uparrow-\varrho_\downarrow\equiv\Delta\varrho_{\uparrow\uparrow},\label{3.2}\\
\frac{4}{\cal V}\sum_{\bf p}f_{33}({\bf
p})&=(\varrho_{n\uparrow}+\varrho_{p\downarrow})-
(\varrho_{n\downarrow}+\varrho_{p\uparrow})\equiv\Delta\varrho_{\uparrow\downarrow}.\label{3.4}
 \end{align}
 Here $\alpha$ is the isospin asymmetry parameter, $\varrho_{n\uparrow},\varrho_{n\downarrow}$ and
 $\varrho_{p\uparrow},\varrho_{p\downarrow}$ are the neutron and
 proton number densities with spin up and spin down,
 respectively;
 $\varrho_\uparrow=\varrho_{n\uparrow}+\varrho_{p\uparrow}$ and
$\varrho_\downarrow=\varrho_{n\downarrow}+\varrho_{p\downarrow}$
are the nucleon densities with spin up and spin down. The
quantities $\Delta\varrho_{\uparrow\uparrow}$ and
$\Delta\varrho_{\uparrow\downarrow}$ may be regarded as FM and AFM
spin order parameters. Indeed, in symmetric nuclear matter, if all
nucleon spins are aligned in one direction (totally polarized FM
spin state), then $\Delta\varrho_{\uparrow\uparrow}=\varrho$ and
$\Delta\varrho_{\uparrow\downarrow}=0$; if all neutron spins are
aligned in one direction and  all proton spins in the opposite one
(totally polarized  AFM spin state), then
$\Delta\varrho_{\uparrow\downarrow}=\varrho$ and
$\Delta\varrho_{\uparrow\uparrow}=0$. In turn, from
Eqs.~\p{3.1}--\p{3.4} one can find the
neutron and proton number densities with spin up and spin down as functions of
the total density $\varrho$, isospin excess $\delta\varrho\equiv
\alpha\varrho$, and FM and AFM order parameters
$\Delta\varrho_{\uparrow\uparrow}$ and
$\Delta\varrho_{\uparrow\downarrow}$: \bal
\varrho_{n\uparrow}&=\frac{1}{4}(\varrho+\delta\varrho+\Delta\varrho_{\uparrow\uparrow}+
\Delta\varrho_{\uparrow\downarrow}),\nonumber\\
\varrho_{n\downarrow}&=\frac{1}{4}(\varrho+\delta\varrho-\Delta\varrho_{\uparrow\uparrow}-
\Delta\varrho_{\uparrow\downarrow}),\nonumber\\
\varrho_{p\uparrow}&=\frac{1}{4}(\varrho-\delta\varrho+\Delta\varrho_{\uparrow\uparrow}-
\Delta\varrho_{\uparrow\downarrow}),\nonumber\\ 
\varrho_{p\downarrow}&=\frac{1}{4}(\varrho-\delta\varrho-\Delta\varrho_{\uparrow\uparrow}+
\Delta\varrho_{\uparrow\downarrow}).\nonumber\end{align}

In order to characterize spin ordering in the neutron and  proton
subsystems, it is convenient to introduce   neutron and proton
spin polarization parameters \beqe
\Pi_n=\frac{\varrho_{n\uparrow}-\varrho_{n\downarrow}}{\varrho_n},\quad
\Pi_p=\frac{\varrho_{p\uparrow}-\varrho_{p\downarrow}}{\varrho_p}.
\end{equation} The expressions for the spin order parameters
$\Delta\varrho_{\uparrow\uparrow}$ and
$\Delta\varrho_{\uparrow\downarrow}$ through the spin polarization
parameters read \begin{align}
\Delta\varrho_{\uparrow\uparrow}=\varrho_n\Pi_n+\varrho_p\Pi_p,\;
\Delta\varrho_{\uparrow\downarrow}=\varrho_n\Pi_n-\varrho_p\Pi_p.\nonumber
\end{align}

 To obtain the self--consistent equations, we specify the energy functional of
the system in the form
\begin{align} E(f)&=E_0(f)+E_{int}(f), \label{14}\\
{E}_0(f)&=4\sum\limits_{ \bf p}^{} \varepsilon_0({\bf
p})f_{00}({\bf p}),\;\varepsilon_0({\bf p})=\frac{{\bf
p}^{\,2}}{2m_{0}},
\\ {E}_{int}(f)&=2\sum\limits_{ \bf p}^{}\{
\tilde\varepsilon_{00}({\bf p})f_{00}({\bf p})+
\tilde\varepsilon_{30}({\bf p})f_{30}({\bf p})\label{13.1}\\
&\quad+\tilde\varepsilon_{03}({\bf p})f_{03}({\bf p})+
\tilde\varepsilon_{33}({\bf p})f_{33}({\bf p})\} ,
\nonumber\end{align} \begin{align}\tilde\varepsilon_{00}({\bf
p})&=\frac{1}{2\cal V}\sum_{\bf q}U_0({\bf k})f_{00}({\bf
q}),\;{\bf k}=\frac{{\bf p}-{\bf q}}{2}, \nonumber\\
\tilde\varepsilon_{30}({\bf p})&=\frac{1}{2\cal V}\sum_{\bf
q}U_1({\bf k})f_{30}({\bf q}),\nonumber\\ 
\tilde\varepsilon_{03}({\bf p})&=\frac{1}{2\cal V}\sum_{\bf
q}U_2({\bf k})f_{03}({\bf q}), \nonumber\\
\tilde\varepsilon_{33}({\bf p})&=\frac{1}{2\cal V}\sum_{\bf
q}U_3({\bf k})f_{33}({\bf q}). \nonumber
\end{align}
 Here
  $m_0$ is the bare mass of a nucleon, $U_0({\bf k}),...,U_3({\bf k}) $ are the normal FL
amplitudes, and
$\tilde\varepsilon_{00},\tilde\varepsilon_{30},\tilde\varepsilon_{03},\tilde\varepsilon_{33}$
are the FL corrections to the free single particle spectrum.
Further we do not take into account the effective tensor forces,
which lead to coupling of the momentum and spin degrees of freedom
\cite{HJ,D,FMS}, and, correspondingly, to anisotropy in the
momentum dependence of FL amplitudes with respect to the spin
polarization axis. Using Eqs.~\p{1} and \p{14}, we
get the self--consistent equations in the form \bal \xi_{00}({\bf
p})&=\varepsilon_{0}({\bf p})+\tilde\varepsilon_{00}({\bf
p})-\mu_{00}^0,\;
\xi_{30}({\bf p})=\tilde\varepsilon_{30}({\bf p}), \\
\xi_{03}({\bf p})&=\tilde\varepsilon_{03}({\bf p})-\mu_{03}^0, \;
\xi_{33}({\bf p})=\tilde\varepsilon_{33}({\bf p}).
\nonumber\end{align}
  To obtain
 numerical results, we  use the Skyrme effective interaction.
In the case of Skyrme forces the normal FL amplitudes
read
 \begin{align}
U_0({\bf k})&=6t_0+t_3\varrho^\beta
+\frac{2}{\hbar^2}[3t_1+t_2(5+4x_2)]{\bf k}^{2},\label{13}
\\
U_1({\bf
k})&=-2t_0(1-2x_0)-\frac{1}{3}t_3\varrho^\beta(1-2x_3)\nonumber\\
&\quad-\frac{2}{\hbar^2}[t_1(1-2x_1)-t_2(1+2x_2) ]{\bf
k}^{2}\equiv a+b{\bf k}^{2},
\nonumber\\
U_2({\bf
k})&=-2t_0(1+2x_0)-\frac{1}{3}t_3\varrho^\beta(1+2x_3)\nonumber\\
&\quad-\frac{2}{\hbar^2}[t_1(1+2x_1)- t_2(1+2x_2)]{\bf k}^{2},\nonumber\\
U_3({\bf k})&=-2t_0-\frac{1}{3}t_3\varrho^\beta
-\frac{2}{\hbar^2}(t_1- t_2){\bf k}^{2}\equiv c+d{\bf
k}^{2},\nonumber
\end{align}
where $t_i,x_i,\beta$ are the phenomenological constants,
characterizing a given parametrization of the Skyrme forces.
Eqs.~\p{13} are derived in Appendix. In the  numerical
calculations we
  shall use   SkM$^*$~\cite{BGH} and SGII~\cite{SG} potentials,
  designed for describing the properties of systems with small
  isospin asymmetry, and
SLy4 and SLy5 potentials~\cite{CBH},
  developed to fit the properties of systems with large isospin
  asymmetry.
 With
account of the evident form of FL amplitudes and
Eqs.~\p{3.1}--\p{3.4}, one can obtain \begin{align}
\xi_{00}&=\frac{p^2}{2m_{00}}-\mu_{00}, \\
\xi_{03}&=\frac{p^2}{2m_{03}}-\mu_{03},\\
\xi_{30}&=(a+b\frac{{\bf
p}^{2}}{4})\frac{\Delta\varrho_{\uparrow\uparrow}}{8}+\frac{b}{32}\langle
{\bf q}^{2}\rangle_{30}, \label{4.3}\\
\xi_{33}&=(c+d\frac{{\bf
p}^{2}}{4})\frac{\Delta\varrho_{\uparrow\downarrow}}{8}+\frac{d}{32}\langle
{\bf q}^{2}\rangle_{33},\label{4.4}
\end{align}
where the effective nucleon mass $m_{00}$ and  effective isovector
mass
  $m_{03}$ are defined by
 the formulae:
\begin{eqnarray}
\frac{\hbar^2}{2m_{00}}&=&\frac{\hbar^2}{2m_0}+\frac{\varrho}{16}
[3t_1+t_2(5+4x_2)],\label{18}\\
\frac{\hbar^2}{2m_{03}}&=& \frac{\alpha\varrho}{16}[t_2(1+2x_2)-
t_1(1+2x_1)],\nonumber\end{eqnarray} and the renormalized chemical
potentials $\mu_{00}$ and $\mu_{03}$ should be determined from
Eqs. \p{3.1} and \p{3.3}. In Eqs. ~\p{4.3} and  \p{4.4}, $\langle
{\bf q}^{2}\rangle_{30}$ and $\langle {\bf q}^{2}\rangle_{33}$ are
the second order moments of the corresponding distribution
functions
\begin{align} \langle {\bf
q}^{2}\rangle_{30}&=\frac{4}{V}\sum_{\bf q}{\bf
q}^2f_{30}({\bf q}),\label{6.1}\\
\langle {\bf q}^{2}\rangle_{33}&=\frac{4}{V}\sum_{\bf q}{\bf
q}^2f_{33}({\bf q}). \label{6.2}\end{align}

Thus, with account of  expressions \p{2.4} for the distribution
functions $f$, we obtain the self--consistent equations
\p{3.1}--\p{3.4}, \p{6.1}, and \p{6.2} for the effective chemical
potentials $\mu_{00},\mu_{03}$,
 FM  and AFM spin
 order parameters
$\Delta\varrho_{\uparrow\uparrow}$,
$\Delta\varrho_{\uparrow\downarrow}$, and  second order moments
$\langle {\bf q}^{2}\rangle_{30}, \langle {\bf q}^{2}\rangle_{33}$.
It is easy to see, that the self--consistent equations remain
invariable under a global flip of  spins, when neutrons (protons)
with spin up and spin down are interchanged, and under a global
flip of  isospins, when   neutrons and protons with the same spin
projection are interchanged.

Let us consider, what differences will be in the case of neutron
matter.  Neutron matter is an infinite nuclear system, consisting
of nucleons of one species, i.e., neutrons, and, hence, the
formalism of one--component Fermi liquid should be applied for the
description of its properties. Formally neutron matter can be
considered as the limiting case of asymmetric nuclear matter,
corresponding to the isospin asymmetry $\alpha=1$. The individual
state of a neutron is  characterized by momentum $\bf p$ and spin
projection $\sigma$. The self--consistent equation has the form of
Eq.~\p{2}, where all quantities are matrices in the space of
$\kappa\equiv(\tbf p,\sigma)$ variables. The normal distribution
function and single particle energy can be expanded in the Pauli
matrices in spin space \bal  f({\bf p})&= f_{0}({\bf
p})\sigma_0+f_{3}({\bf p})\sigma_3,
\label{7.23}\\
  \varepsilon({\bf p})&= \varepsilon_{0}({\bf p})\sigma_0+\varepsilon_{3}({\bf
p})\sigma_3. \nonumber
\end{align}

The energy functional of neutron matter is characterized by two
normal FL amplitudes $U_0^n(\tbf k)$ and $U_1^n(\tbf k)$.
 The normal FL amplitudes
can be found in terms of the Skyrme force parameters
$t_i,x_i,\beta$ (the details of the derivation procedure are given
in Appendix):
\bal U_0^n({\bf k})&=2t_0(1-x_0)+\frac{t_3}{3}\varrho^\beta(1-x_3)\label{101}\\&\quad
+\frac{2}{\hbar^2}[t_1(1-x_1)+3t_2(1+x_2)]{\bf k}^{2},
\nonumber\\
U_1^n({\bf
k})&=-2t_0(1-x_0)-\frac{t_3}{3}\varrho^\beta(1-x_3)\label{102}\\&\quad
+\frac{2}{\hbar^2}[t_2(1+x_2)-t_1(1-x_1)]{\bf k}^{2}\equiv
a_n+b_n{\bf k}^{2}.\nonumber\end{align} With account of
Eqs.~\p{101} and \p{102}, the normalization conditions for the
distribution functions can be written in the form
\bal\frac{2}{\cal V}\sum_{\bf p}f_{0}({\bf p})&=\varrho,\;
\label{3.17}\\
\frac{2}{\cal V}\sum_{\bf p}f_{3}({\bf
p})&=\varrho_\uparrow-\varrho_\downarrow\equiv\Delta\varrho_{\uparrow\uparrow}.
\label{3.30}
\end{align}
Here  $\varrho_\uparrow$ and $\varrho_\downarrow$ are the neutron
number densities with spin up and spin down and
\begin{align}
f_{0}&=\frac{1}{2}\{n(\omega_{+})+n(\omega_{-})
\},\quad\omega_\pm=\xi_0\pm\xi_3,\label{4.38}\\
f_{3}&=\frac{1}{2}\{n(\omega_{+})-n(\omega_{-})\},\label{4.39}\\
\xi_{0}&=\frac{p^2}{2m_{n}}-\mu_{n},\\
\xi_{3}&=(a_n+b_n\frac{{\bf
p}^{2}}{4})\frac{\Delta\varrho_{\uparrow\uparrow}}{4}+\frac{b_n}{16}\langle
{\bf q}^{2}\rangle_{3}. \label{4.33}
\end{align}
The effective neutron mass $m_{n}$  is defined by
 the formula
\begin{eqnarray}
\frac{\hbar^2}{2m_{n}}&=&\frac{\hbar^2}{2m_0}+\frac{\varrho}{8}
[t_1(1-x_1)+3t_2(1+x_2)],\label{181}\end{eqnarray} and the
quantity $\langle {\bf q}^{2}\rangle_{3}$ in Eq.~\p{4.33}  is the
second order moment of the distribution function $f_3$:
\begin{align} \langle {\bf
q}^{2}\rangle_{3}&=\frac{2}{V}\sum_{\bf q}{\bf q}^2f_{3}({\bf
q}).\label{6.11}\end{align}  Thus, with account of  expressions
\p{4.38} and  \p{4.39} for the distribution functions $f$, we
obtain the self--consistent equations \p{3.17}, \p{3.30}, and
\p{6.11} for the effective chemical potential $\mu_{n}$,
  spin  order parameter
$\Delta\varrho_{\uparrow\uparrow}$,
 and  second order moment
$\langle {\bf q}^{2}\rangle_{3}$. 
\section{Phase transitions in symmetric nuclear matter}
Early research on spin polarizability  of nuclear matter with the
Skyrme effective interaction were based on the calculation of
magnetic susceptibility and finding  its pole
structure~\cite{VNB,RPLP}, determining the onset of   instability
with respect to spin fluctuations. Here we shall find directly
solutions of the self--consistent equations for the FM and AFM
spin order parameters as functions of density at zero temperature.
In this section a special emphasis will be laid on the study of
symmetric nuclear matter ($\alpha=0$), while in the next section
on the investigation of strongly asymmetric nuclear matter
($\alpha\lesssim1$), including neutron matter as its limiting
case.
\begin{figure}[b]
\includegraphics[height=12.6cm,width=8.6cm,trim=49mm 105mm 56mm 46mm,
draft=false,clip]{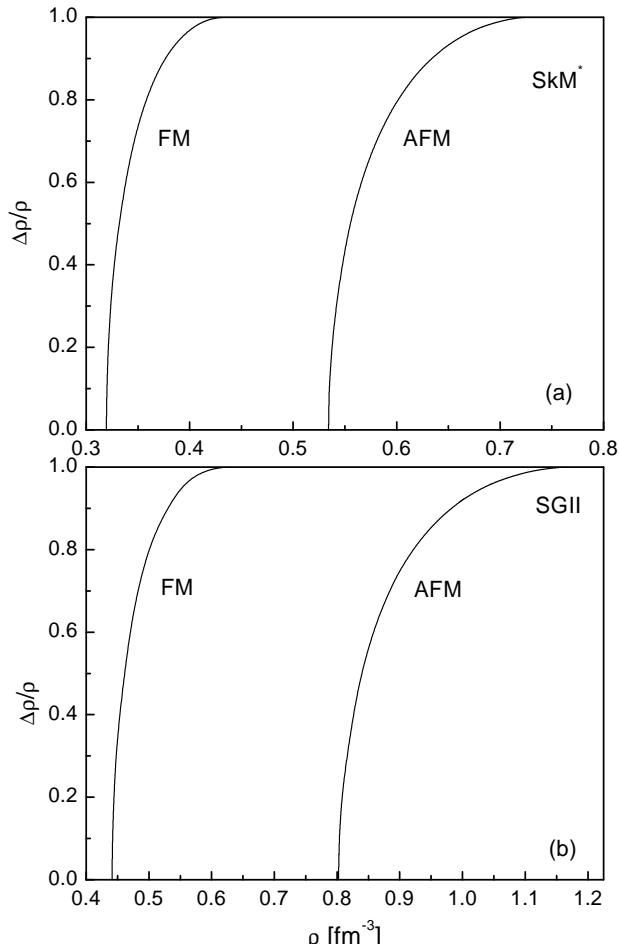} \caption{FM and AFM spin
polarization parameters as functions of density at zero
temperature for (a) SkM$^*$ force and (b) SGII force.
}\label{fig1}
\end{figure}

Let us consider the zero temperature behavior of spin polarization
in symmetric nuclear matter ($\varrho_p=\varrho_n$). The FM spin
ordering corresponds to the case
$\Delta\varrho_{\uparrow\uparrow}\not=0,\langle {\bf
q}^{2}\rangle_{30}\not=0,\Delta\varrho_{\uparrow\downarrow}=0,\langle
{\bf q}^{2}\rangle_{33}=0$, while the AFM spin ordering  to the
case $\Delta\varrho_{\uparrow\downarrow}\not=0,\langle {\bf
q}^{2}\rangle_{33}\not=0,\Delta\varrho_{\uparrow\uparrow}=0,\langle
{\bf q}^{2}\rangle_{30}=0$. In the totally ferromagnetically
polarized state nontrivial solutions of the self--consistent
equations have the form \begin{equation}
\Delta\varrho_{\uparrow\uparrow}=\varrho,\; \langle {\bf
q}^{2}\rangle_{30}=\frac{3}{5}\varrho
k_F^2.\label{18.1}\end{equation} Here $k_F=(3\pi^2\varrho)^{1/3}$
is Fermi momentum  of symmetric nuclear matter in the case when
degrees of freedom, corresponding to spin up of nucleons, are open
while those related to spin down are inaccessible. For the totally
antiferromagnetically polarized nuclear matter we have
\begin{equation} \Delta\varrho_{\uparrow\downarrow}=\varrho,\;
\langle {\bf q}^{2}\rangle_{33}=\frac{3}{5}\varrho
k_F^2.\end{equation} The Fermi momentum $k_F$ is given by the same
expression as in Eq.~\p{18.1} since now degrees of freedom,
related to spin down of protons and spin up of neutrons, are
inaccessible.
 The results of numerical
determination of FM $\Delta\varrho_{\uparrow\uparrow}/\varrho$ and
AFM $\Delta\varrho_{\uparrow\downarrow}/\varrho$ spin polarization
parameters are shown in Fig.~\ref{fig1}  for the SkM$^*$ and SGII
effective forces.

The FM spin order parameter arises at density
$\varrho\approx2\varrho_0$ for the SkM$^*$ potential and at
$\varrho\approx2.75\varrho_0$ for the SGII potential. The AFM
order parameter originates at $\varrho\approx3.3\varrho_0$ for the
SkM$^*$ force and at $\varrho\approx5\varrho_0$ for the SGII
force. In both cases FM ordering appears earlier than AFM one.
Nuclear matter becomes totally ferromagnetically polarized
($\Delta\varrho_{\uparrow\uparrow}/\varrho=1$) at density
$\varrho\approx2.7\varrho_0$ for the SkM$^*$ force and at
$\varrho\approx3.9\varrho_0$ for the SGII force. Totally
antiferromagnetically polarized state
($\Delta\varrho_{\uparrow\downarrow}/\varrho=1$) is formed at
$\varrho\approx4.5\varrho_0$ for the SkM$^*$ potential and at
$\varrho\approx7.2\varrho_0$ for the SGII potential.

Note that in symmetric nuclear matter the neutron and proton spin
polarization parameters for the FM spin ordering are given by the
formulas \beqe \Pi_n=\Pi_p=\frac{\Delta
\varrho_{\uparrow\uparrow}}{\varrho}\nonumber\end{equation} and for the
AFM spin ordering their expressions read \beqe
\Pi_n=-\Pi_p=\frac{\Delta
\varrho_{\uparrow\downarrow}}{\varrho}.\nonumber\end{equation}

\begin{figure}[t]
\includegraphics[height=12.6cm,width=8.6cm,trim=49mm 103mm 56mm 46mm,
draft=false,clip]{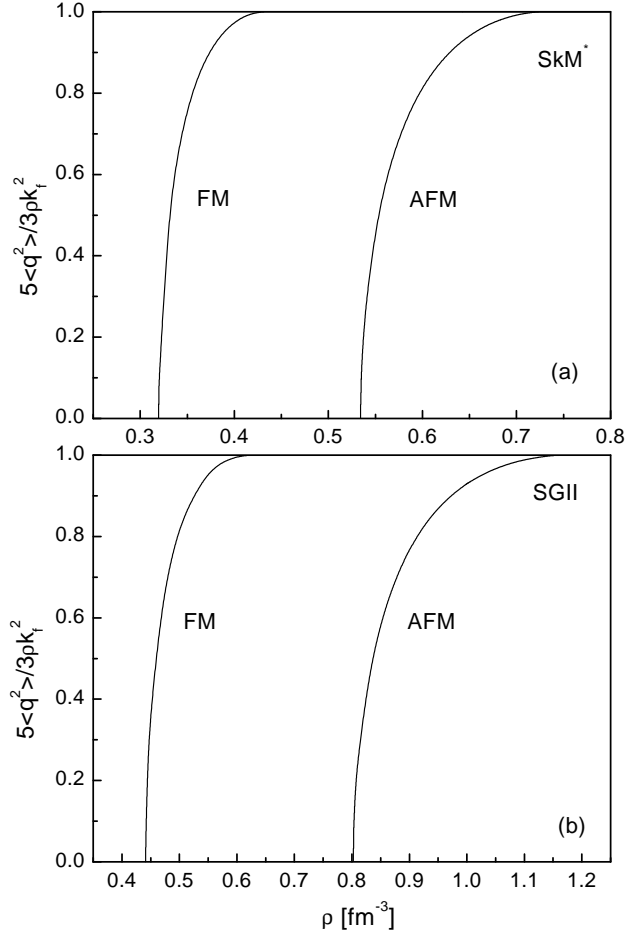} \caption{Same as in Fig.~\ref{fig1},
but for the second order moments of the distribution functions.
}\label{fig2}
\end{figure}

The second order moments $\langle {\bf q}^{2}\rangle_{30}, \langle
{\bf q}^{2}\rangle_{33}$ of the distribution functions
$f_{30},f_{33}$ also play the role of the order parameters. In
Fig.~\ref{fig2} it is shown behavior of these quantities
normalized to their  value in the totally polarized state. The
ratios $5\langle {\bf q}^{2}\rangle_{30}/3\varrho k_F^2$ and
$5\langle {\bf q}^{2}\rangle_{33}/3\varrho k_F^2$ are regarded as
FM and AFM order parameters, respectively.
The behavior of these quantities is similar to that of the spin
polarization parameters in Fig.~\ref{fig1}, with the same values
of the threshold densities for the appearance and saturation of
the order parameters.

\begin{figure}[tb]
\includegraphics[height=12.6cm,width=8.6cm,trim=49mm 103mm 56mm 46mm,
draft=false,clip]{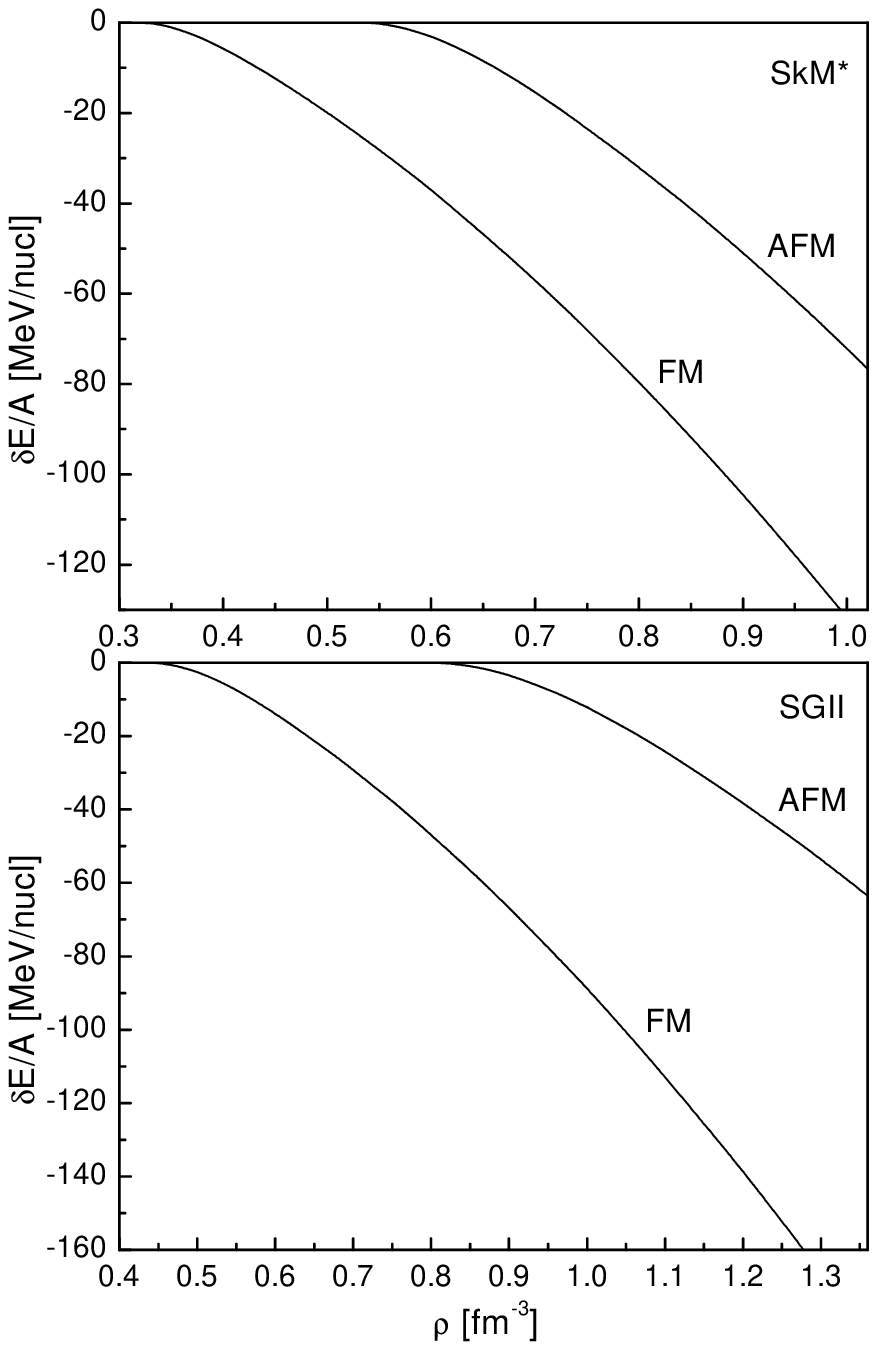} \caption{ Total energy   per
nucleon, measured from its value in the normal state, for FM and
AFM spin ordering  as a function of density at zero temperature
for (a) SkM$^*$ force and (b) SGII force. }\label{fig3}
\end{figure}

 In the density domain, where FM and AFM solutions of
self--consistent equations exist simultaneously, it is necessary
to clarify, which solution is thermodynamically preferable. To
this end,  we should compare the free energies of both states. The
results of the numerical calculation of the total energy   per
nucleon, measured from its value in the normal state, are shown in
Fig.~\ref{fig3}.
One can see, that for all relevant densities FM spin ordering is
more preferable than AFM one, and, moreover, the difference
between corresponding free energies becomes larger with increasing
density, so that there is no evidence, that AFM spin ordering
could become preferable at larger densities. These results are in
correspondence with the results of the Brueckner calculations with
Nijmegen NSC97e potential in Ref.~\cite{VB}, where it was shown
that for symmetric nuclear matter the state with the oppositely
directed spins of protons and neutrons is less favorable than the
state with the same direction of nucleon spins for all relevant
densities. However, in contrast to calculations in Ref.~\cite{VB},
we find that for high density region  there will be realized the
FM spin ordering of nucleon spins as a ground state of symmetric
nuclear matter instead of the nonpolarized state.
\section{Phase transitions in strongly asymmetric nuclear matter}

In this section we continue to study the properties of spin
polarized nuclear matter, but now in the region of large isospin
asymmetry. In contrast to symmetric nuclear matter, the analysis
shows that in strongly asymmetric nuclear matter it is realized
the antiparallel spin ordering of neutron and proton spins.

 If all
neutron and proton spins are aligned in one direction, then for
nontrivial solutions of the self--consistent equations we have
\bal \Delta\varrho_{\uparrow\uparrow}&=\varrho,\quad
\Delta\varrho_{\uparrow\downarrow}=\alpha\varrho,\\
\langle {\bf q}^{2}\rangle_{30}&=\frac{3}{10}\varrho
k_F^2[(1+\alpha)^{5/3}+(1-\alpha)^{5/3}],\nonumber\\ \langle {\bf
q}^{2}\rangle_{33}&=\frac{3}{10}\varrho
k_F^2[(1+\alpha)^{5/3}-(1-\alpha)^{5/3}],\nonumber\end{align}
where
 $k_F=(3\pi^2\varrho)^{1/3}$
is the Fermi momentum  of totally polarized symmetric nuclear
matter. Therefore, for the partial number densities of nucleons
with spin up and spin down one can get
\begin{equation}
\varrho_{n\uparrow}=\frac{1+\alpha}{2}\varrho,\;
\varrho_{p\uparrow}=\frac{1-\alpha}{2}\varrho,\;
\varrho_{p\downarrow}=\varrho_{n\downarrow}=0.\end{equation}

If   all neutron spins are aligned in one direction and  all
proton spins in the opposite one, then  \bal
\Delta\varrho_{\uparrow\uparrow}&=\alpha\varrho,\quad
\Delta\varrho_{\uparrow\downarrow}=\varrho,\\
\langle {\bf q}^{2}\rangle_{30}&=\frac{3}{10}\varrho
k_F^2[(1+\alpha)^{5/3}-(1-\alpha)^{5/3}],\nonumber\\ \langle {\bf
q}^{2}\rangle_{33}&=\frac{3}{10}\varrho
k_F^2[(1+\alpha)^{5/3}+(1-\alpha)^{5/3}],\nonumber\end{align} and,
hence,
\begin{equation}
\varrho_{n\uparrow}=\frac{1+\alpha}{2}\varrho,\;
\varrho_{p\downarrow}=\frac{1-\alpha}{2}\varrho,\;
\varrho_{p\uparrow}=\varrho_{n\downarrow}=0.\end{equation}

\begin{figure}[t]
\includegraphics[height=12.6cm,width=8.6cm,trim=49mm 105mm 56mm 46mm,
draft=false,clip]{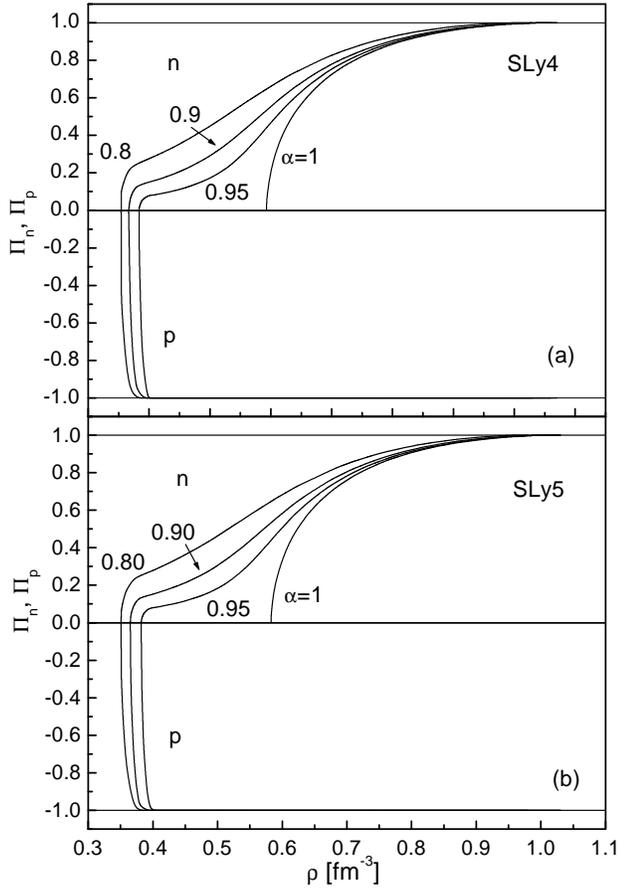} \caption{Neutron and proton spin
polarization parameters as  functions of density  at zero
temperature for (a) SLy4 force and (b) SLy5 force. }\label{fig4}
\end{figure}

Now we present the results of the numerical solution of the
self--consistent equations with the effective SLy4 and SLy5 forces
for strongly asymmetric nuclear ($\alpha=0.95, 0.9, 0.8$) and
neutron ($\alpha=1$) matter.      The neutron and proton spin
polarization parameters $\Pi_n$ and $\Pi_p$  are shown in
Fig.~\ref{fig4} as
 functions of density at zero temperature. Since in a
polarized state the signs of spin polarizations are opposite,
considering solutions correspond to the case, when spins of
neutrons and protons are aligned in the opposite direction. Note
that for SLy4 and SLy5 forces, being relevant for the description
of strongly asymmetric nuclear matter, there are no solutions
corresponding to the same direction of neutron and proton spins.
The reason is that the sign of the multiplier $t_3(-1+2x_3)$ in
the density dependent term of the FL amplitude $U_1$, determining
spin--spin correlations, is positive, and, hence, corresponding
term increases with the increase of nuclear matter density,
preventing instability with respect to spin fluctuations.
Contrarily, the density dependent term $-t_3\varrho^\beta/3$ in
the FL amplitude $U_3$, describing spin--isospin correlations, is
negative, leading to spin instability with the oppositely directed
spins of neutrons and protons at higher densities.
\begin{figure}[tb]
\includegraphics[height=12.6cm,width=8.6cm,trim=49mm 103mm 56mm 46mm,
draft=false,clip]{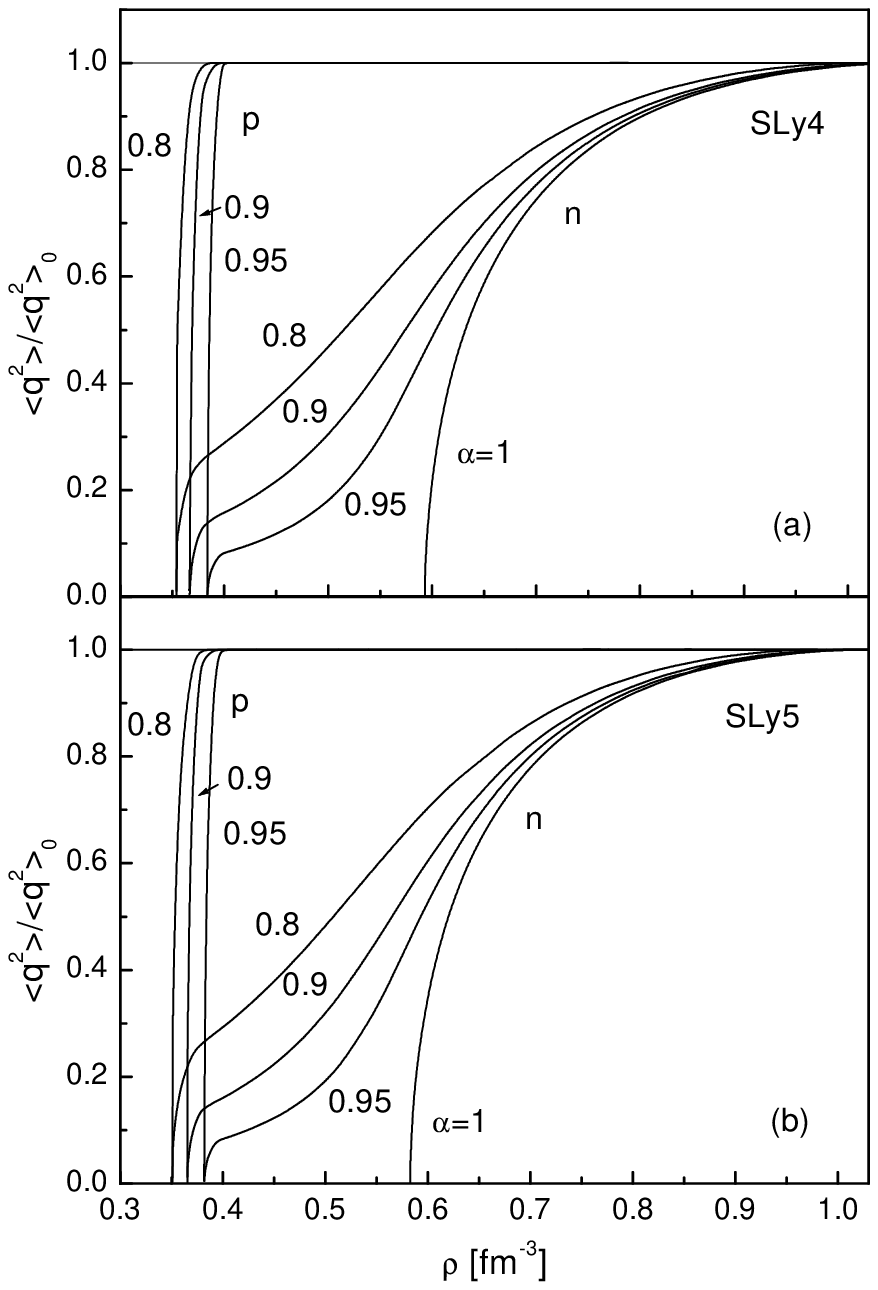} \caption{Same as in Fig.~\ref{fig4},
but for the second order moments $\langle {\bf q}^{2}\rangle_{p}$
and $\langle {\bf q}^{2}\rangle_{n}$,  normalized to their values
in the totally polarized state. }\label{fig5}
\end{figure}

Another nontrivial feature relates to the density behavior of the
spin polarization parameters at large isospin asymmetry. As seen
from Fig.~\ref{fig4}, even small admixture of protons leads to the
appearance of long tails in the density profiles of the neutron
spin polarization parameter near the transition point to a spin
ordered state. As a consequence, the spin polarized state is
formed much earlier in density than in pure neutron matter. For
example, the critical density in neutron matter is
$\varrho\approx0.59\,\mbox{fm}^{-3}$ for SLy4 potential and
$\varrho\approx0.58\,\mbox{fm}^{-3}$  for SLy5 potential; in
asymmetric nuclear matter with $\alpha=0.95$ the spin polarized
state arises at $\varrho\approx0.38\,\mbox{fm}^{-3}$ for SLy4
 and  SLy5 potentials. Hence, even small
quantity of protons strongly favors spin instability of highly
asymmetric nuclear matter, leading to the appearance of states
with the oppositely directed spins of  neutrons and protons. As
follows from Fig.~\ref{fig4}, protons become totally spin
polarized within a very narrow density domain (e.g., if
$\alpha=0.95$, full polarization occurs at
$\varrho\approx0.41\,\mbox{fm}^{-3}$ for SLy4 force and at
$\varrho\approx0.40\,\mbox{fm}^{-3}$ for SLy5 force) while the
threshold densities for the appearance and saturation of the
neutron spin order parameter are substantially different (if
$\alpha=0.95$, neutrons become totally polarized at
$\varrho\approx1.05\,\mbox{fm}^{-3}$ for SLy4 force and  at
$\varrho\approx1.02\,\mbox{fm}^{-3}$ for SLy5 force).

Note that the second order moments \bal \langle {\bf
q}^{2}\rangle_{n}&\equiv\langle {\bf
q}^{2}\rangle_{n\uparrow}-\langle {\bf
q}^{2}\rangle_{n\downarrow}\\
 &=\frac{1}{V}\sum_{\bf q} {\bf
q}^2\Bigl(n(\omega_{+,+})-n(\omega_{-,+})\Bigr),\nonumber\\
\langle {\bf q}^{2}\rangle_{p}&\equiv\langle {\bf
q}^{2}\rangle_{p\uparrow}-\langle {\bf
q}^{2}\rangle_{p\downarrow}\nonumber\\
 &=\frac{1}{V}\sum_{\bf q}
{\bf q}^2\Bigl(n(\omega_{+,-})-n(\omega_{-,-})\Bigr)
\nonumber\end{align}
 also characterize spin polarization of the
neutron and proton subsystems. If the solutions $\langle {\bf
q}^{2}\rangle_{30}$ and $\langle {\bf q}^{2}\rangle_{33}$ of the
self--consistent equations are known, then \bal \langle {\bf
q}^{2}\rangle_{n}&=\frac{1}{2}\bigl(\langle {\bf
q}^{2}\rangle_{30}+\langle {\bf q}^{2}\rangle_{33}\bigr),\nonumber\\
\langle {\bf q}^{2}\rangle_{p}&=\frac{1}{2}\bigl(\langle {\bf
q}^{2}\rangle_{30}-\langle {\bf
q}^{2}\rangle_{33}\bigr).\nonumber\end{align}
\begin{figure}[t]
\includegraphics[height=12.6cm,width=8.6cm,trim=49mm 103mm 56mm 46mm,
draft=false,clip]{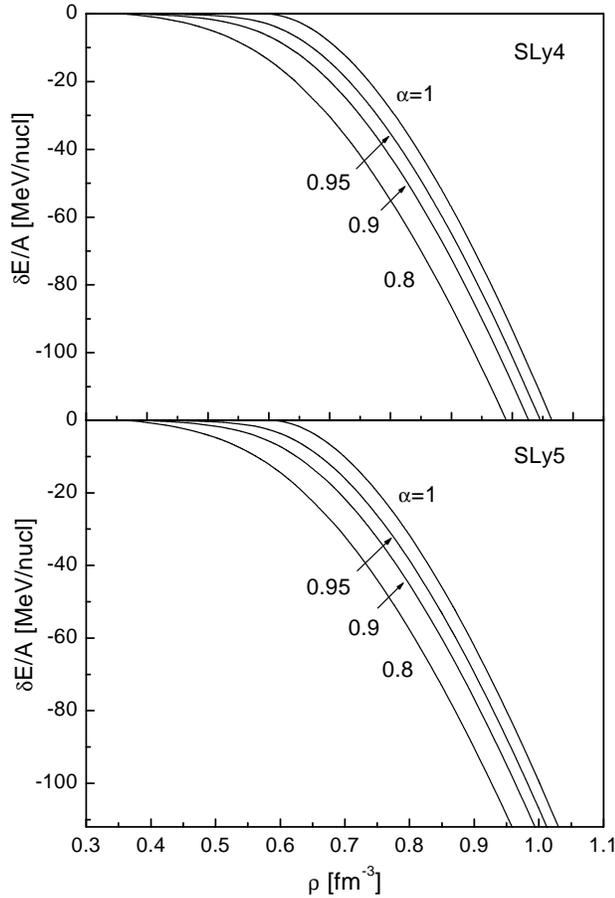} \caption{Total energy   per nucleon,
measured from its value in the normal state,  for the state with
the oppositely directed spins of neutrons and protons as a
function of density at zero temperature for (a) SLy4 force and (b)
SLy5 force. }\label{fig6}
\end{figure}
The values of $\langle {\bf q}^{2}\rangle_{n}$ and $\langle {\bf
q}^{2}\rangle_{p}$ for the totally polarized state are \beqestar
\langle {\bf q}^{2}\rangle_{n0}=\frac{3}{10}\varrho
k_F^2(1+\alpha)^{5/3},\;\langle {\bf
q}^{2}\rangle_{p0}=-\frac{3}{10}\varrho
k_F^2(1-\alpha)^{5/3}.\end{equation*}

In Fig.~\ref{fig5} we plot the density dependence of the second
order moments $\langle {\bf q}^{2}\rangle_{n}$ and $\langle {\bf
q}^{2}\rangle_{p}$, normalized to their values in the totally
polarized state, for different asymmetries at zero temperature.
These quantities behave similarly to the spin polarization
parameters in Fig.~\ref{fig4}, i.e., there exist long tails in the
density profiles of the  neutron spin order parameter and the
proton spin order parameter is saturated within a very narrow
density interval.

To check thermodynamic stability of the spin ordered state with
the oppositely directed spins of neutrons and protons, it is
necessary to compare the free energies of this state and the
normal state. In Fig.~\ref{fig6}  the difference between  the
total energies per nucleon  of the spin ordered and normal states
is shown as a function of density at zero temperature. One can see
that nuclear matter undergoes a phase transition to the state with
the oppositely directed  spins of neutrons and protons at some
critical density, depending on the isospin asymmetry.

Note that our results of neutron matter calculations obtained with
the Skyrme effective interaction predict a FM phase transition at
some critical density, that is different from the results of
calculations with realistic NN potentials~\cite{VPR,FSS}. In
Ref.~\cite{VPR}, employing Nijmegen II and Reid93 NN potentials,
it has been found within the Brueckner--Hartree--Fock approach,
that in the range of densities explored
($\varrho\lesssim7\varrho_0$) totally polarized neutron matter is
always more repulsive than nonpolarized matter, being an
indication that a phase transition of neutron matter to a FM state
is not expected. Besides, in Ref.~\cite{FSS}, in connection with
the problem of the neutrino diffusion in dense matter, it has been
shown within the framework of the auxiliary field diffusion Monte
Carlo method with the Argonne $v_{18}$ two--body potential and
Urbana IX three--body potential, that the magnetic susceptibility
of neutron matter shows a strong reduction of about a factor 3
with respect to its Fermi gas value. However, calculations of the
magnetic susceptibility with the Skyrme effective
interaction~\cite{VNB} give indication of infinite discontinuity,
and, hence, predict a FM transition at densities
$\varrho\approx0.18-0.26\,\mbox{fm}^{-3}$.

For strongly asymmetric nuclear matter with Skyrme forces we find
a phase transition to the state with antiparallel spins of
neutrons and protons, that is different from the results of
calculations with NSC97e NN potential in Ref.~\cite{VB}, where the
nonpolarized  state was predicted for the whole range of densities
up to 1.2 $\mbox{fm}^{-3}$. The reasons, explaining such
disagreement in calculations with the effective and realistic NN
potentials, are discussed in the next section.

\section{Discussion and Conclusions}

Spin instability is a common feature, associated with a large
class of Skyrme models, but is not realized in more microscopic
calculations. The Skyrme interaction has been successful in
describing nuclei and their excited states. In addition, various
authors have exploited its applicability to describe bulk matter
at densities of relevance to neutron stars \cite{SMK}. The force
parameters are determined empirically by calculating the ground
state in the Hartree--Fock approximation and by fitting the
observed ground state properties of nuclei and nuclear matter at
the saturation density. In particular, the interaction parameters,
describing spin--spin and spin--isospin correlations, are
constrained from the data on isoscalar \cite{T,LS} and isovector
(giant Gamow--Teller)~\cite{HHR,SGE,BDE} spin--flip resonances.

In a microscopic approach, one starts with the bare interaction
and obtains an effective particle--hole interaction by solving
iteratively the Bethe--Goldstone equation. In contrast to the
Skyrme models, calculations with realistic NN potentials predict
more repulsive total energy per particle for a polarized state
comparing to the  nonpolarized one for all relevant densities,
and, hence, give no  indication of a phase transition to a spin
ordered state. It must be emphasized that the interaction in the
spin-- and isospin--dependent channels is a crucial ingredient in
calculating spin properties of isospin symmetric and asymmetric
nuclear matter and different behavior at high densities of the
interaction amplitudes, describing spin--spin and spin--isospin
correlations, lays behind this divergence in calculations with the
effective and realistic potentials.

In this study as a potential of NN interaction we chose SkM$^*$
and SGII (symmetric nuclear matter) as well as  SLy4 and SLy5
(strongly asymmetric nuclear matter) Skyrme effective forces. The
models SkM$^*$ and SGII~\cite{SG} have been constrained by fitting
the properties of nucleon systems with very small isospin
asymmetries, while the models  SLy4 and SLy5 were further
constrained to reproduce the results of microscopic neutron matter
calculations (pressure versus density curve)~\cite{CBH}. Besides,
in a recent publication~\cite{SMK} it was shown that the density
dependence of the nuclear symmetry energy, calculated up to
densities $\varrho\lesssim3\varrho_0$ with SLy4 and SLy5 effective
forces (together with some other sets of parameters among the
total 87 Skyrme force parametrizations checked) gives the neutron
star models in a broad agreement with the observables, such as the
minimum rotation period, gravitational mass--radius relation, the
binding energy, released in supernova collapse, etc. This is an
important check for using these parametrizations in the high
density region of strongly asymmetric nuclear matter. However, it
is necessary to note, that the spin--dependent part of the Skyrme
interaction at densities of relevance to neutron stars still
remains  to be constrained. Probably, these constraints will be
obtained from the data on the time decay of magnetic field of
isolated neutron stars~\cite{PP}. In spite of this shortcoming,
SLy4 and SLy5 effective forces hold one of the most competing
Skyrme parametrizations at present time for description of isospin
asymmetric nuclear matter at high density (together with SkM$^*$
and SGII forces for description of symmetric nuclear matter) while
a Fermi liquid approach with Skyrme effective forces provides a
consistent and transparent framework for studying spin
instabilities in a nucleon system.

 In summary, we have considered the
possibility of phase transitions into spin ordered states of
symmetric and strongly asymmetric nuclear matter within the Fermi
liquid formalism, where NN interaction is described by the Skyrme
effective forces (SkM$^*$, SGII and SLy4, SLy5 potentials for the
regions of vanishing and strong isospin asymmetry, respectively).
In contrast to the previous considerations, where the possibility
of formation of FM spin polarized states was studied on the base
of calculation of magnetic susceptibility,  we obtain the
self--consistent equations for the FM and AFM spin order
parameters and solve them in the case of zero temperature. It has
been found that nuclear matter demonstrates different behavior at
high densities with respect to spin fluctuations in isospin
symmetric and strongly isospin asymmetric cases. In the model with
SkM$^*$ and SGII effective forces symmetric nuclear matter
undergoes a FM phase transition, when the spins of protons and
neutrons are aligned along the same direction. In the model with
SLy4 and SLy5 effective forces strongly asymmetric nuclear matter
is subjected to a phase transition into the spin polarized state
with the oppositely directed spins of neutrons and protons, while
the state with the same direction of the neutron and proton spins
does not appear. In the last case, an important peculiarity of the
corresponding phase transition is the existence of long tails in
the density profile of the neutron spin polarization parameter
near the transition point. This means that even small admixture of
protons to neutron matter leads to a considerable shift of the
critical density of spin instability in the direction of low
densities. In the model with SLy4 effective interaction this
displacement is from the critical density
$\varrho\approx3.7\varrho_0$ for neutron matter to
$\varrho\approx2.4\varrho_0$ for asymmetric nuclear matter at the
isospin asymmetry $\alpha=0.95$, i.e. for $2.5\%$ of protons only.
As a result,  the state with the oppositely directed spins of
neutrons and protons appears, where protons become totally
polarized in a very narrow density domain.

As a consequence of this study,  important questions appear, what
is the value of the threshold asymmetry, at which the parallel
spin ordering at small isospin asymmetry is changed to the
antiparallel spin ordering at large isospin asymmetry, and do the
obtained results survive for another type of an effective
interaction, e.g., for Gogny effective force~\cite{BGG,FVS} or
monopole  effective interaction~\cite{RSM}? This research is in
progress and will be reported elsewhere.

\section*{Appendix}
The aim of this section is to establish relationships between the
Fermi liquid amplitudes and amplitudes of NN interaction in the
leading order on the interaction between nucleons. To this end we
present the Hamiltonian of the system in the form \bal
H&=H_0+V,\label{39}\\
H_0=\sum_{\kappa_1\kappa_2}\varepsilon^0_{\kappa_1\kappa_2}a^+_{\kappa_1}a_{\kappa_2},\quad
V&=\frac{1}{2\cal V}\sum_{\kappa_1\kappa_2\kappa_3\kappa_4}\hat
v(\kappa_1\kappa_2;\kappa_3\kappa_4)a^+_{\kappa_1}a^+_{\kappa_2}a_{\kappa_4}a_{\kappa_3},\nonumber
\end{align}
where $\hat v$ is the amplitude of NN interaction. We shall assume
that the amplitude $\hat v$ doesn't depend on the total momentum
of colliding nucleons, but only from their relative momenta:\bal
\hat v(\kappa_1\kappa_2;\kappa_3\kappa_4)&=\hat v({\bf p}_1,{\bf
p}_2;{\bf
p}_3,{\bf p}_4)_{\underline\kappa_1\underline\kappa_2;\underline\kappa_3\underline\kappa_4},
\quad \underline\kappa\equiv(\sigma,\tau),\\
\hat v({\bf p}_1,{\bf p}_2;{\bf p}_3,{\bf
p}_4)_{\underline\kappa_1\underline\kappa_2;\underline\kappa_3\underline\kappa_4}&=\hat
v({\bf p},{\bf
q})_{\underline\kappa_1\underline\kappa_2;\underline\kappa_3\underline\kappa_4};
\quad {\bf p}=\frac{{\bf p}_1 -{\bf p}_2}{2},\; {\bf q}=\frac{{\bf
p}_3 -{\bf p}_4}{2}.\nonumber
\end{align}
To obtain the energy functional, corresponding to the
Hamiltonian~\p{39}, we should average the Hamiltonian~\p{39} over
the state of  nonideal gas of particles. In the leading
approximation on the interaction, using the Wick rules and taking
into account that $\mbox{Tr}\varrho
a^+_{\kappa_2}a_{\kappa_1}\equiv f_{\kappa_1\kappa_2}$ ($\varrho$
being the statistical operator), one
gets \bal E(f)&=E_0(f)+E_{int}(f)\label{41}\\
E_0(f)=\sum_{\kappa_1\kappa_2}\varepsilon^0_{\kappa_1\kappa_2}f_{\kappa_2\kappa_1},\quad
E_{int}(f)&=\frac{1}{2\cal
V}\sum_{\kappa_1\kappa_2\kappa_3\kappa_4} \hat
v(\kappa_1\kappa_2;\kappa_3\kappa_4)\bigl(f_{\kappa_3\kappa_1}f_{\kappa_4\kappa_2}-
f_{\kappa_4\kappa_1}f_{\kappa_3\kappa_2}\bigr). \nonumber
\end{align}
Hence, according to Eq.~\p{1}, expression for the single particle
energy reads \bal
\varepsilon_{\kappa_1\kappa_2}&=\varepsilon_{\kappa_1\kappa_2}^0(f)+\tilde
\varepsilon_{\kappa_1\kappa_2}(f),\label{42}\\
\tilde\varepsilon_{\kappa_1\kappa_2}(f)=\frac{1}{2\cal
V}\sum_{\kappa_3\kappa_4}\bigl\{ \hat
v(\kappa_1\kappa_3;\kappa_2\kappa_4)&+\hat
v(\kappa_3\kappa_1;\kappa_4\kappa_2)- \hat
v(\kappa_1\kappa_3;\kappa_4\kappa_2)-\hat
v(\kappa_3\kappa_1;\kappa_2\kappa_4)
\bigr\}f_{\kappa_4\kappa_3}.\nonumber\end{align} For spatially
homogeneous distributions \bal
f_{\kappa_1\kappa_2}=f_{\underline\kappa_1\underline\kappa_2}({\bf
p}_1)\delta_{p_1,p_2},\quad
\varepsilon_{\kappa_1\kappa_2}=\varepsilon_{\underline\kappa_1\underline\kappa_2}({\bf
p_1})\delta_{p_1,p_2}\end{align} and Eqs.~\p{41}, \p{42} can be
simplified \bal
E_0(f)&=\sum_{\underline\kappa_1,\underline\kappa_2;\,\bf p
}\varepsilon^0_{\underline\kappa_1\underline\kappa_2}({\bf
p}) f_{\underline\kappa_2\underline\kappa_1}({\bf p}),\label{44}\\
 E_{int}(f)&=\frac{1}{2}\sum_{\underline\kappa_1,\underline\kappa_2;\,{\bf
p}} \tilde \varepsilon_
{\underline\kappa_1\underline\kappa_2}({\bf
p})f_{\underline\kappa_2\underline\kappa_1}({\bf p}), \nonumber
\end{align}
where \bal \tilde\varepsilon_
{\underline\kappa_1\underline\kappa_2}({\bf p})&= \frac{1}{2\cal
V}\sum_{\underline\kappa_3\underline\kappa_4}\bigl\{ \hat
v_{\underline\kappa_1\underline\kappa_3;\underline\kappa_2\underline\kappa_4}({\bf
k},{\bf k})+\hat
v_{\underline\kappa_3\underline\kappa_1;\underline\kappa_4\underline\kappa_2}(-{\bf
k},-{\bf k})\label{45}\\&\quad - \hat
v_{\underline\kappa_1\underline\kappa_3;\underline\kappa_4\underline\kappa_2}({\bf
k},-{\bf k})- \hat
v_{\underline\kappa_3\underline\kappa_1;\underline\kappa_2\underline\kappa_4}(-{\bf
k},{\bf k})\bigr\} f_{\underline\kappa_4\underline\kappa_3}({\bf
q}),\quad {\bf k}=\frac{{\bf p}-{\bf q}}{2}. \nonumber \end{align}
The general structure of the NN interaction amplitude $\hat v$,
invariant with respect to rotations in spin and isospin spaces,
has the form \beq \hat v({\bf p},{\bf q})=v_0({\bf p},{\bf q})+
v_1({\bf p},{\bf q}) {{\boldsymbol\sigma_1\boldsymbol\sigma_2}}+
v_2({\bf p},{\bf q}) {{\boldsymbol\tau_1\boldsymbol\tau_2}} +
v_3({\bf p},{\bf q})(
{{\boldsymbol\sigma_1\boldsymbol\sigma_2}})({{\boldsymbol\tau_1\boldsymbol\tau_2}}).
\eeq Taking into account the general structure of the normal
distribution function \beq f({\bf p})=f_{00}({\bf p})+f_{k0}({\bf
p})\sigma_k+f_{0l}({\bf p})\tau_l+f_{kl}({\bf
p})\sigma_k\tau_l\eeq and calculating traces in Eqs.~\p{44},
\p{45}, one can get \bal {E}_0(f)&=4\sum\limits_{ \bf p}^{}
\varepsilon_0({\bf p})f_{00}({\bf p}),\;\varepsilon_0({\bf
p})=\frac{{\bf
p}^{\,2}}{2m_{0}},
\\ {E}_{int}(f)&=2\sum\limits_{ \bf p}^{}\{
\tilde\varepsilon_{00}({\bf p})f_{00}({\bf p})+
\tilde\varepsilon_{k0}({\bf p})f_{k0}({\bf p})\label{51}\\
&\quad+\tilde\varepsilon_{0l}({\bf p})f_{0l}({\bf p})+
\tilde\varepsilon_{kl}({\bf p})f_{kl}({\bf p})\} ,
\nonumber\end{align} \begin{align}\tilde\varepsilon_{00}({\bf
p})&=\frac{1}{2\cal V}\sum_{\bf q}U_0({\bf k})f_{00}({\bf
q}),\;{\bf k}=\frac{{\bf p}-{\bf q}}{2}, \nonumber\\
\tilde\varepsilon_{k0}({\bf p})&=\frac{1}{2\cal V}\sum_{\bf
q}U_1({\bf k})f_{k0}({\bf q}),\nonumber\\ 
\tilde\varepsilon_{0l}({\bf p})&=\frac{1}{2\cal V}\sum_{\bf
q}U_2({\bf k})f_{0l}({\bf q}), \nonumber\\
\tilde\varepsilon_{kl}({\bf p})&=\frac{1}{2\cal V}\sum_{\bf
q}U_3({\bf k})f_{kl}({\bf q}), \nonumber
\end{align}
where \bal U_0({\bf k})&=\bigl\{4v_0({\bf k},{\bf k})-v_0(-{\bf
k},{\bf k})-3v_1(-{\bf k},{\bf k})-3v_2(-{\bf k},{\bf k})-
9v_3(-{\bf k},{\bf k})\bigr\}+\bigl\{{\bf k}\rightarrow-{\bf k}\bigr\},\nonumber\\
U_1({\bf k})&=\bigl\{4v_1({\bf k},{\bf k})-v_0(-{\bf k},{\bf
k})+v_1(-{\bf k},{\bf k})-3v_2(-{\bf k},{\bf k})+
3v_3(-{\bf k},{\bf k})\bigr\}+\bigl\{{\bf k}\rightarrow-{\bf k}\bigr\},\label{48}\\
U_2({\bf k})&=\bigl\{4v_2({\bf k},{\bf k})-v_0(-{\bf k},{\bf
k})-3v_1(-{\bf k},{\bf k})+v_2(-{\bf k},{\bf k})+ 3v_3(-{\bf
k},{\bf k})\bigr\}+\bigl\{{\bf k}\rightarrow-{\bf k}\bigr\},\nonumber\\
U_3({\bf k})&=\bigl\{4v_3({\bf k},{\bf k})-v_0(-{\bf k},{\bf
k})+v_1(-{\bf k},{\bf k})+v_2(-{\bf k},{\bf k})- v_3(-{\bf k},{\bf
k})\bigr\}+\bigl\{{\bf k}\rightarrow-{\bf k}\bigr\}.\nonumber
\end{align}
The interaction energy functional~\p{13.1} represents a special
case of the functional~\p{51}, corresponding to a collinear spin
ordering. The amplitude of NN interaction for the Skyrme effective
forces has the form \bal\hat v({\bf p},{\bf
q})&=t_0(1+x_0P_\sigma)+\frac{1}{6}t_3(1+x_3P_\sigma)\varrho^\beta+\frac{1}{2\hbar^2}
t_1(1+x_1P_\sigma)({\bf p}^2+{\bf q}^2)\label{49}\\ &\quad
+\frac{t_2}{\hbar^2}(1+x_2P_\sigma){\bf p}{\bf q},\quad
P_\sigma=\frac{1}{2}(1+{{\boldsymbol\sigma_1\boldsymbol\sigma_2}}),\nonumber\end{align}
Here $P_\sigma$ is the spin exchange operator. Extracting from
Eq.~\p{49} the functions $v_0,...,v_3$ and substituting their
expressions into Eq.~\p{48}, we obtain Eqs.~\p{13} for the normal
FL amplitudes $U_0,..., U_3$, describing, respectively, density,
spin, isospin and spin--isospin correlations in a nucleon Fermi
liquid.

For neutron matter the basic equations~\p{41}, \p{42} remain
unchanged, but now the individual state of a neutron is
characterized by momentum $\bf p$ and spin projection $\sigma$,
$\kappa\equiv({\bf p},\sigma)$. Besides, taking into account the
general structure of the interaction amplitude $\hat v$ and the
normal distribution function $f$ \bal \hat v({\bf p},{\bf
q})&=v_0({\bf p},{\bf q})+ v_1({\bf p},{\bf q})
{{\boldsymbol\sigma_1\boldsymbol\sigma_2}},\\
f({\bf p})&=f_{0}({\bf p})+f_{k}({\bf p})\sigma_k,
 \end{align}
and calculating traces with the Pauli matrices $\sigma_i$, we
obtain \bal {E}_0(f)&=2\sum\limits_{ \bf p}^{} \varepsilon_0({\bf
p})f_{0}({\bf p}),\;\varepsilon_0({\bf p})=\frac{{\bf
p}^{\,2}}{2m_{0}},\nonumber
\\ {E}_{int}(f)&=\sum\limits_{ \bf p}^{}\{
\tilde\varepsilon_{0}({\bf p})f_{0}({\bf p})+
\tilde\varepsilon_{k}({\bf p})f_{k}({\bf p})\},
\nonumber\end{align} where
\begin{align}\tilde\varepsilon_{0}({\bf p})&=\frac{1}{2\cal
V}\sum_{\bf q}U_0^n({\bf k})f_{0}({\bf
q}),\;{\bf k}=\frac{{\bf p}-{\bf q}}{2}, \\
\tilde\varepsilon_{k}({\bf p})&=\frac{1}{2\cal V}\sum_{\bf
q}U_1^n({\bf k})f_{k}({\bf q}),\nonumber
\end{align}
and the FL amplitudes are given by the formulas \bal U_0^n({\bf
k})&=\bigl\{2v_0({\bf k},{\bf k})-v_0(-{\bf k},{\bf
k})-3v_1(-{\bf k},{\bf k})\bigr\}+\bigl\{{\bf k}\rightarrow-{\bf k}\bigr\},\label{53}\\
U_1^n({\bf k})&=\bigl\{2v_1({\bf k},{\bf k})-v_0(-{\bf k},{\bf
k})+v_1(-{\bf k},{\bf k})\bigr\}+\bigl\{{\bf k}\rightarrow-{\bf
k}\bigr\}\label{}.\nonumber
\end{align}
After substituting expressions for the functions $v_0,v_1$ from
Eq.~\p{49}, we obtain Eqs.~\p{101}, \p{102} for the neutron matter
FL amplitudes $U_0^n, U_1^n$.

\end{document}